\documentclass
[preprint,byrevtex,nofootinbib,10pt,twocolumn,balancelastpage,titlepage,noshowkeys,bibnotes]{revtex4}%
\usepackage{amsfonts}
\usepackage{amsmath}
\usepackage{amssymb}
\usepackage{graphicx}%
\providecommand{\U}[1]{\protect\rule{.1in}{.1in}}
\ifx\pdfoutput\relax\let\pdfoutput=\undefined\fi
\newcount\msipdfoutput
\ifx\pdfoutput\undefined\else
\ifcase\pdfoutput\else
\ifx\paperwidth\undefined\else
\ifdim\paperheight=0pt\relax\else\pdfpageheight\paperheight\fi
\ifdim\paperwidth=0pt\relax\else\pdfpagewidth\paperwidth\fi
\fi\fi\fi
\begin{document}
\preprint{ }
\preprint{UATP/2204}
\title{Overlooked Work and Heat of Intervention and the Fate of Information
Principles of Szilard and Landauer}
\author{P.D. Gujrati,}
\affiliation{$^{1}$Department of Physics, $^{2}$School of Polymer Science and Engineering,
The University of Akron, Akron, OH 44325}
\email{pdg@uakron.edu}

\begin{abstract}
We show that any external intervention (insertion or removal of a partition)
that destroys the equilibrium or brings it in a system always requires work
and heat to ensure that the first law is obeyed, a fact that has been
completely overlooked in the literature. As a consequence, there is no second
law violation. We discuss the ramifications of our finding for information
principles of Szilard and Landauer and show that no information entropy is
needed. The relevance of this result for Maxwell's demon is also considered.

\end{abstract}
\date{\today}
\maketitle

\begin{figure}
[ptb]
\begin{center}
\includegraphics[
height=1.6362in,
width=3.8346in
]%
{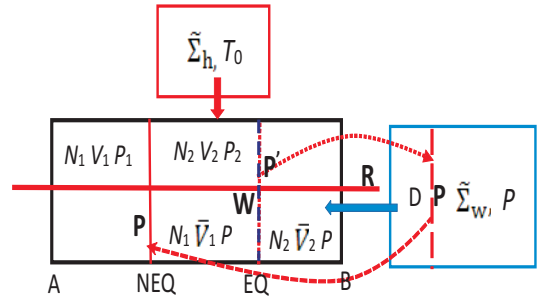}%
\caption{The system $\Sigma$ of ideal gas with $N$ particles in a volume $V$
with pressure $P$ and temperature $T_{0}$\ kept constant by a heat medium
$\widetilde{\Sigma}_{\text{h}}$. The demon D is part of a work medium
$\widetilde{\Sigma}_{\text{w}}$ and manipulates the partition \textbf{P} for
intervention by inserting it at NEQ (which divides $\Sigma$ into $\Sigma_{1}$
on its left and $\Sigma_{2}$ on its right) and removing from EQ as described
in the text, the latter also defines an imaginary wall \textbf{W}; see the
text. The rod \textbf{R} and \textbf{P} move \emph{together} when connected.
The rod \textbf{R} is used to identify the state from outside the system after
intervention. }%
\label{Fig-Intervention}%
\end{center}
\end{figure}

Szilard is said to be the first one to exorcise Maxwell's demon \cite{Maxwell}
by introducing the concept of the "intervention of intelligent beings" (the
demon D) in a milestone paper \cite{Szilard} dealing with what is
conventionally called a Maxwell's pressure-demon\ system \cite{Rex}. The
system $\Sigma$ is an ideal gas with $N$ particles in thermal contact with a
heat bath $\widetilde{\Sigma}_{\text{h}}$ at fixed temperature $T_{0}$ (see
the red arrow representing thermal contact in Fig. \ref{Fig-Intervention}),
which D\ separates into volumes $V_{\text{1}}$\ and $V_{\text{2}}$\ by
\emph{inserting} a partition \textbf{P} at a fixed location NEQ (see the long
dashed red arrow); the partition, which is \emph{rigid} and \emph{impervious}
to particle flows across it, can also be used as a piston. The demon and
\textbf{P} shown by a broken red vertical line forms a part of a mechanical
work source $\widetilde{\Sigma}_{\text{w}}$ \cite{Note}; see the blue arrow
for mechanical contact. The idea is to distinguish four distinct processes (i)
\emph{intervention} during a period $\Delta\tau_{\text{I}}$ by inserting or
removing the partition \emph{without} requiring any work or heat, (ii)
\emph{informational measurement} of determining the state of the system during
$\Delta\tau_{\text{M}}$, (iii) \emph{utilization of measurement }by extracting
work from the gas at the expense of heat extraction from $\widetilde{\Sigma
}_{\text{h}}$ by isothermal expansion during $\Delta\tau_{\text{E}}$, which is
the only work considered, and (iv) removal of \textbf{P} from EQ and return it
into $\widetilde{\Sigma}_{\text{w}}$. The demon also passes (removes) a rod
\textbf{R} through $\Sigma$ and \textbf{P} as shown when inserting (removing)
\textbf{P.} The engine extracts a positive work in a cycle, and violates the
second law. Szilard \emph{postulates} that the information in (ii)\ is
correlated with "... a certain definite average entropy production..." $\Delta
S_{\text{SL}}$\ (SL for Szilard-Landauer) to salvage the second law. Using
this, Brillouin \cite{Brillouin0} proclaims that every physical measurement
requires minimum entropy increase to be performed.%
%TCIMACRO{\FRAME{ftbpFU}{3.8346in}{1.6362in}{0pt}{\Qcb{The system $\Sigma$ of
%ideal gas with $N$ particles in a volume $V$ with pressure $P$ and temperature
%$T_{0}$\ kept constant by a heat medium $\widetilde{\Sigma}_{\text{h}}$. The
%demon D is part of a work medium $\widetilde{\Sigma}_{\text{w}}$ and
%manipulates the partition \QTR{bf}{P} for intervention by inserting it at NEQ
%(which divides $\Sigma$ into $\Sigma_{1}$ on its left and $\Sigma_{2}$ on its
%right) and removing from EQ as described in the text, the latter also defines
%an imaginary wall \QTR{bf}{W}; see the text. The rod \QTR{bf}{R} and
%\QTR{bf}{P} move \QTR{em}{together} when connected. The rod \QTR{bf}{R} is
%used to identify the state from outside the system after intervention.
%}}{\Qlb{Fig-Intervention}}{intervention0.EPS}%
%width 3.8346in;  height 1.6362in;  depth 0pt;  original-width 6.5561in;
%original-height 2.0678in;  cropleft "0.0423";  croptop "1.0373";
%cropright "0.8163";  cropbottom "0";
%filename 'Intervention0.EPS';file-properties "XNPEU";}} }%
%BeginExpansion
%EndExpansion
\ 

Landauer argues that $\Delta S_{\text{SL}}$ is instead due to resetting
\cite{Landauer,Plenio,Frank,Lent,Klaers}. Chambadal \cite{Chambadal} later
argues that entropy and information are not involved in (i) and (ii). Penrose
\cite{Penrose} emphasizes fluctuations to generate work and create a perpetual
machine. These attempts have inspired a highly active informational approach
to thermodynamics that is still undergoing rapid growth
\cite{Mallick,Avery,Wolpert}.\ 

Although Szilard's treatment is completely classical, there are recent quantum
mechanicanical treatments; see for example \cite{Zurek,Ueda,Serra}. However,
there are still contentious issues in treating information as thermodynamic
entropy and not all agree with the significance of information for entropy and
work. A nice discussion of these issues and various suggestions is given by
Leff and Rex \cite{Rex}. It has been pointed out \cite{Kish} that relocation
of the partition (not the insertion itself) must be carefully accounted to
avoid any violation of the second law or that a deeper connection exists
between thermodynamic efficiency and irrelevant information \cite{Still}.
Thus, the controversy and confusion persist \cite{Norton,Bub,Bennett0}.
Despite this, information principles of Szilard and Landauer have been
visionary in creating the modern field of cybernetics and artificial
intelligence in which information is processed by machines.

Among all the confusion, the omission of the work and heat of intervention
until now seems most disturbing as they most probably contribute not only to
the net work and heat in a complete cycle but are also necessarily required by
the first law. Unfortunately, this law has not received any attention to date
as the entire field has been mired by the need to satisfy the second law
alone. Indeed to the best of our knowledge, almost all researchers including
Szilard and Landauer believe that classically, the intervention neither
involves heat nor work. This is quite surprising as any possible entropy
change $\Delta S_{\text{I}}$ due to intervention results in a directly
measurable heat transfer $\Delta Q_{\text{I}}$ that plays an important role,
and requires some accompanying work $\Delta W_{\text{I}}$ to satisfy the first
law. But $\Delta S_{\text{I}}$ must be \emph{nonzero} if the intervention
desturbs the state by destroying the equilibrium state $\mathfrak{M}%
_{\text{eq}}$\ or bringing it back. Then it must play a pivotal role in not
only settling the violation of the second law but also satisfying the first
law. This also suggests that almost all earlier attempts to satisfy the second
law will need reassessment.

Therefore, there is an urgent need to carefully look at the process of
intervention in (i) and the role of information in (ii). We discuss a
many-particle system to make a general argument about intervention for an
arbitrary system. Our main conclusion is in the form of the following theorem
that intervention by itself usually requires nonzero $\Delta S_{\text{I}}$,
$\Delta Q_{\text{I}}$, and $\Delta W_{\text{I}}$, which so far have been
overlooked, and have profound implications for the underpinnings of the entire
field of information theory.

\textbf{Intervention Theorem}: \textit{Any act of intervention such as
insertion (Szilard) or removal (Landauer) of a partition that destroys
}$M_{\text{eq}}$\textit{ or brings it back in an interacting system always
results in heat that can be externally measured and satisfies the first law to
determine work.}

\textbf{Proof}: We consider a prefixed point NEQ that divides $\Sigma$ into
fixed volumes $V_{1}$ on its left and $V_{2}$ on its right. Initially, the gas
is in $\mathfrak{M}_{\text{eq}}$. At time $t=0$, D moves \textbf{P} from
$\widetilde{\Sigma}_{\text{w}}$\ and inserts it at NEQ (different from EQ)
along with \textbf{R}. The partition divides $N$ \emph{randomly} into
$N_{1}\geq0$ and $N_{2}=N-N_{1}\geq0$. Different attempts to put \textbf{P} at
NEQ will result in different values of $N_{1}$ and $N_{2}$. The two gases,
each in equilibrium at some time $\tau_{\text{I}}$, are not in equilibrium
with each other; this state is denoted by $\mathfrak{M}_{\text{neq}}$. We give
two independent methods to find $\Delta W_{\text{I}}$.

(a) Let $\Delta S_{\text{I}}=S(\mathfrak{M}_{\text{neq}})-S(\mathfrak{M}%
_{\text{eq}})$ be the entropy change. Associated with this is the externally
measured heat $\Delta Q_{\text{I}}$. From the first law applied to the
intervention in (i), the corresponding work is $\Delta W_{\text{I}%
}=dQ_{\text{I}}-dE_{\text{I}}$.

(b) We consider the imaginary wall \textbf{W} at EQ in $\mathfrak{M}%
_{\text{eq}}$ so that $N_{1}$ particles are in the volume $\bar{V}_{1}$ to its
left and $N_{2}$ particles in $\bar{V}_{2}$ to its right, and $P$ is the
pressure on both sides. The entropy is $S(\mathfrak{M}_{\text{eq}})$. We treat
\textbf{W} as an imaginary piston \textbf{P}$^{\prime}$ (\textbf{R }is not
important in this process) that moves reversibly from EQ to NEQ so that
$\bar{V}_{1}\rightarrow V_{1}$, and simultaneously $\bar{V}_{2}\rightarrow
V_{2}$ to reproduce $\mathfrak{M}_{\text{neq}}$ in (a). As $\mathfrak{M}%
_{\text{neq}}$ has been reproduced, it is irrelevant how it is generated as
$\Delta S_{\text{I}}$ and $\Delta E_{\text{I}}$ are the same. If
\textbf{P}$^{\prime}$ is moved so that $\Delta Q_{\text{I}}$ is the same as in
(a), then $\Delta W_{\text{I}}$ is also the same as above. As the two gases in
$\bar{V}_{1}$ and $\bar{V}_{2}$\ are in equilibrium with each other, D can
replace \textbf{W} by inserting \textbf{P }(see the reverse of the dotted
short red arrow now pointing to EQ) there. During this insertion at EQ,
$\Delta W_{\text{I}}^{\prime}=\Delta Q_{\text{I}}^{\prime}=0$. Now, D can use
\textbf{P }in place of \textbf{P}$^{\prime}$\ to move it to NEQ to yield the
same $\Delta W_{\text{I}},\Delta S_{\text{I}}$, and $\Delta Q_{\text{I}}$.

During removal of \textbf{P} from NEQ, D changes $\mathfrak{M}_{\text{neq}}$
to $\mathfrak{M}_{\text{eq}}$ so that all $\Delta S_{\text{I}},\Delta
Q_{\text{I}}$, and $\Delta W_{\text{I}}$\ are negative of the above quantities
for insertion. This proves the theorem.

Before proceeding further, we briefly use the above new understanding of the
insertion of \textbf{P} in a one-particle Szilard engine (Fig.
\ref{Fig-Intervention} with $N_{1}=N=1,N_{2}=0,V_{1}=V_{2}=V/2$). Before
insertion, the probabilities are $p_{1}=p_{2}=1/2$ of the particle being in
$\Sigma_{1}$\ and $\Sigma_{2}$, respectively. After insertion, we have either
$p_{1}=1,p_{2}=0$\ (state $0$ or $\mathfrak{M}_{10}$)\ or $p_{1}=0,p_{2}=1$
(state $1$ or $\mathfrak{M}_{01}$). This results in $\Delta S_{\text{I}}%
=-\ln2\Rightarrow\Delta Q_{\text{I}}=-T_{0}\ln2$ (externally measurable) due
to insertion alone in (i) so $\Delta W_{\text{I}}=-T_{0}\ln2$ from $\Delta
E_{\text{I}}=\Delta Q_{\text{I}}-\Delta W_{\text{I}}=$ $0$. Thus, $\left\vert
\Delta Q_{\text{I}}\right\vert $ is given out to $\widetilde{\Sigma}%
_{\text{h}}$ and $\left\vert \Delta W_{\text{I}}\right\vert $ to
$\widetilde{\Sigma}_{\text{w}}$, \textit{i.e.}, D. All this is in accordance
with the above theorem. This first law for the intervention in (i) has never
been used to date to draw this conclusion.

\textbf{Work and Heat of Intervention}: The above proof is for a general
system but we now determine these quantities for an ideal gas that is assumed
to always have the same temperature $T_{0}$. The pressure $P$ in
$\mathfrak{M}_{\text{eq}}$ satisfies $PV=NT_{0}$. The pressure of the two
gases in $\mathfrak{M}_{\text{neq}}$ at time $\tau_{\text{I}}$ are
$P_{1}=N_{1}T_{0}/V_{1}$ and $P_{2}=N_{2}T_{0}/V_{2}$, respectively. We now
show that
\begin{equation}
\Delta W_{\text{I}}=T_{0}[N\ln P-N_{1}\ln P_{1}-N_{2}\ln P_{2}]\leq0.
\label{Intervention-Work}%
\end{equation}

We first consider (a). Initially, the entropy is $S(\mathfrak{M}_{\text{eq}%
})=N[g(T_{0})-\ln P]$, where $g(T_{0})$ is a constant function of constant
$T_{0}$ \cite{Landau} and plays no role in our discussion so we will omit it
from now on. Moreover, to simplify the discussion, we will restrict ourselves
to reversible processes. The entropy after insertion is given by
$S(\mathfrak{M}_{\text{neq}})=-N_{1}\ln P_{1}-N_{2}\ln P_{2}$. Thus,%
\begin{equation}
\Delta S_{\text{I}}=N\ln P-N_{1}\ln P_{1}-N_{2}\ln P_{2}\leq0,
\label{Intervention-Entropy}%
\end{equation}
the equality (no entropy reduction) occurs if $\Sigma$ is in equilibrium
($\Delta P=P_{1}-P_{2}=0$), which happens when NEQ and\ EQ are the same point.
From $\Delta S_{\text{I}}$, we obtain $\Delta Q_{\text{I}}=T_{0}\Delta
S_{\text{I}}\leq0$ so that $\Delta W_{\text{I}}=\Delta Q_{\text{I}}$ from the
first law, which proves Eq. (\ref{Intervention-Work}). The negative values of
$\Delta Q_{\text{I}}$ and $\Delta W_{\text{I}}$ mean that heat is ejected to
$\widetilde{\Sigma}_{\text{h}}$ and the work is done by D or $\widetilde
{\Sigma}_{\text{w}}$. During the removal of \textbf{P} by D from NEQ, gases
will come to equilibrium with a concomitant entropy gain $\left\vert \Delta
S_{\text{I}}\right\vert $\ so heat $\left\vert \Delta Q_{\text{I}}\right\vert
$\ will be absorbed and work $\left\vert \Delta W_{\text{I}}\right\vert $ done
by the gas on D. We obtain the same values if we follow (b).

The intervention does not require $\Delta S_{\text{I}},\Delta Q_{\text{I}}$,
and $\Delta W_{\text{I}}$ if it does not destroy $\mathfrak{M}_{\text{eq}}$
($\Delta P=0$); otherwise they are nonzero but always satisfy the first law.
We have only considered reversible processes as our goal is to merely justify
their existence. It should be noted that \textbf{P} is impermeable to particle
flow so once $N_{1}$ is determined, it cannot change. In this regard, our
treatment is different from that in the quantum version of Szilard's engine
\cite{Ueda}. Our approach is in keeping with Szilard-Landauer idea that
$\mathfrak{M}_{10}$ and $\mathfrak{M}_{01}$ for $N=1$ are distinct states.

\textbf{Measurement and Information:} We now follow the consequences of the
new understanding of the intervention for the measurement process. We treat
\textbf{P}-\textbf{R} as a movable piston inside $\Sigma$. Before the
intervention, we know $V_{1},V_{2},P,T_{0}$, and $N$. After intervention, we
need to know internal quantities $N_{1},N_{2}$ and $P_{1},P_{2}$ for a
complete thermodynamic description. We now show that thermodynamics alone is
sufficient to determine them so D does not need to \emph{actively} "gather"
any information. The \emph{spontaneous} motion of \textbf{P} is always towards
equilibrium at \textbf{W} during which $\Delta P\rightarrow0$, which D easily
determines by simply observing the direction \textbf{R} moves from the outside
without making any mechanical contact with $\Sigma$ so no work is involved: if
moving to the right, $\Delta P>0$ and $\bar{V}_{1}>V_{1},\bar{V}_{2}<V_{2}$;
if to the left, $\Delta P<0$ and $\bar{V}_{1}<V_{1},\bar{V}_{2}>V_{2}$. As
soon as \textbf{R} moves, D may attach a weight on its opposite end to perform
external work. We will not consider any weight as our interest is to show what
unknown quantities are thermodynamically determined. Observing \textbf{W
}where \textbf{P} stops from the outside, D determines $\bar{V}_{1}$ and
$\bar{V}_{2}$, again without any work involved. The ideal gas equation
gives$\ \ \ \ \ \ \ \ \ $
\ \ \ \ \ \ \ \ \ \ \ \ \ \ \ \ \ \ \ \ \ \ \ \ \ \ \ \ \ \ \ \ \ \ \ \ \ \ \ \ \ \ \ \
\begin{equation}
N_{1}=N\frac{\bar{V}_{1}}{V},N_{2}=N\frac{\bar{V}_{2}}{V},P_{1}=P\frac{\bar
{V}_{1}}{V_{1}},P_{2}=P\frac{\bar{V}_{2}}{V_{2}}.
\label{Values-after-Intervention}%
\end{equation}
Thus, we have verified that the state of the system is completely determined
thermodynamically; there is no need for D to obtain any internal information
that requires any work by mechanical contacts with $\Sigma$. As a consequence,
$\Delta W_{\text{M}}=\Delta Q_{\text{M}}=\Delta E_{\text{M}}=0$. The situation
with Szilard's engine is even simpler; see below.

\textbf{Isothermal Expansion:} The positive work done by the gas in the
isothermal expansion in (iii) from NEQ (\textbf{P)}\ to EQ\ (\textbf{W)} after
insertion is precisely the magnitude $\Delta W_{\text{E}}=\left\vert \Delta
W_{\text{I}}\right\vert $ on $\widetilde{\Sigma}_{\text{w}}$, and the same
amount of positive heat $\Delta Q_{\text{E}}=\left\vert \Delta Q_{\text{I}%
}\right\vert $ is taken from $\widetilde{\Sigma}_{\text{h}}$ (recall that
$\Delta W_{\text{E}}$ is the only work considered by Szilard-Landauer). They
satisfy the first law with $\Delta E_{\text{E}}=0$. As the system is at
equilibrium at \textbf{W,}\ D now takes out \textbf{P} (by separating from
\textbf{R}) from EQ and brings it back into $\widetilde{\Sigma}_{\text{w}}$
without involving any work or heat. This completes the isothermal cycle during
which the net change $\Delta E=\Delta E_{\text{I}}+\Delta E_{\text{M}}+\Delta
E_{\text{E}}=0$. The conclusion is the same if \textbf{P }is removed; see later.

\textbf{Information Entropy: }However, remarkably, by recognizing nonzero
$\Delta W_{\text{I}}=\Delta Q_{\text{I}}$ during the intervention, not only
the net heat $\Delta Q$ taken from $\widetilde{\Sigma}_{\text{h}}$ but also
the net work $\Delta W$ done by the gas on $\widetilde{\Sigma}_{\text{w}}$
vanish during an entire cycle so no change has occurred either in
$\Sigma,\widetilde{\Sigma}_{\text{h}}$, and $\widetilde{\Sigma}_{\text{w}}%
$\ (including D). The second law is not challenged. As the entropy of the
universe has not changed, there is no need for Szilard-Landauer information
conjecture $\Delta S_{\text{SL}}>0$\ due to the measurement. As we have
assumed a reversible cycle, it is not clear why this entropy generation is
needed. It is clear that the physical necessity of $\Delta S_{\text{I}},\Delta
Q_{\text{I}}$, and $\Delta W_{\text{I}}$ ensures that the second law is not
violated so the need for $\Delta S_{\text{SL}}>0$\ loses any physical
significance from the way they have been justified so far.

As no net work is generated per cycle, the process cannot be considered as
forming an engine that can produce any work. It is merely performing a
conventional reversible isothermal (isoenergetic) cycle.

A word of caution. The new value of $N_{1}$ in the state $\mathfrak{M}%
_{N_{1}N_{2}}$ of $\Sigma$ in the new cycle is again a random number and need
not be the previous value, just as in Szilard's engine. We follow the same set
of processes as described above in the previous cycle. Again, there will be no
violation of the second law for the new state $\mathfrak{M}_{N_{1}N_{2}}$.

The case $\mathfrak{M}_{N0}$ ($N_{1}=N,N_{2}=0$) with $P_{1}=PV/V_{1},P_{2}=0$
or $\mathfrak{M}_{0N}$ ($N_{1}=0,N_{2}=N$) with $P_{1}=0,P_{2}=PV/V_{2}$ is
very similar to the case considered by Szilard with the states $\mathfrak{M}%
_{10}$ and $\mathfrak{M}_{01}$. We also see that the case is very exceptional
in that it is one of the most improbable fluctuations in $\Sigma$. In this
particular case, once D sees \textbf{R} move in a certain direction, the state
is immediately known to determine $P_{1}=0,P_{2}$. There is no need to follow
the entire motion of \textbf{P }to EQ. This step is also not needed in
Szilard's engine that we now discuss.

\textbf{Szilard's Engine}: For $N=1$, D lifts \textbf{P} from $\widetilde
{\Sigma}_{\text{w}}$ and inserts it at NEQ to obtain the states $\mathfrak{M}%
_{10}$ and \ $\mathfrak{M}_{01}$. For $\mathfrak{M}_{10}$ ($\mathfrak{M}_{01}%
$) with $P_{1}=PV/V_{1},P_{2}=0$ (with $P_{1}=0,P_{2}=PV/V_{2}$), and
\textbf{W} at the right (left) end of $\Sigma$, we have $\Delta S_{\text{I}%
}=-\ln(V/V_{1}),\Delta Q_{\text{I}}=-T_{0}\ln(V/V_{1})$ and $\Delta
W_{\text{I}}=\Delta Q_{\text{I}}$. The state is uniquely determined by how
\textbf{R} moves; see the discussion of $\mathfrak{M}_{N0}$ and $\mathfrak{M}%
_{0N}$ above. Thus, D knows precisely whether the particle is in
$\mathfrak{M}_{10}$ or \ $\mathfrak{M}_{01}$ without having any mechanical
contact (blue thick arrow) with $\Sigma$\ so $\Delta W_{\text{M}}=0$ due to
this information. As D does not have a thermal contact (red thick arrow), we
also have $\Delta S_{\text{M}}=0$ and $\Delta Q_{\text{M}}=0$, which refutes
the conjecture $\Delta S_{\text{SL}}=\ln(V/V_{1})$ . For the particle in
$\mathfrak{M}_{10}$, \textbf{P} stops at \textbf{W} on the right end of
$\Sigma$ and we have $\Delta S_{\text{E}}=\ln(V/V_{1}),\Delta Q_{\text{E}%
}=T_{0}\ln(V/V_{1})$ and $\Delta W_{\text{E}}=\Delta Q_{\text{E}}$; same for
the particle in $\mathfrak{M}_{10}$. The system is in equilibrium once
\textbf{P} arrives at \textbf{W}, where D removes it and puts it back in
$\widetilde{\Sigma}_{\text{w}}$ for the next cycle without any work and heat.
Obviously, no net work and heat are extracted from the system during each
cycle. Hence, there is neither any violation of the second law nor $\Sigma$
performs as an engine.

\textbf{Landauer's Resetting}: The distinct state $0\doteq\mathfrak{M}_{10}$
or $1\doteq$ $\mathfrak{M}_{01}$ (identified by the motion of \textbf{R}
without any mechanical contact\textbf{)} interact with $\widetilde{\Sigma
}_{\text{h}}$ and $\widetilde{\Sigma}_{\text{w}}$. We take $V_{1}=V_{2}=V/2$.
The demon starts with $0$ and $1$ and always produces the same state, say $0$,
by first removing \textbf{P} from NEQ and putting in $\widetilde{\Sigma
}_{\text{w}}$. As each state is a nonequilibrium state before the
intervention, the removal increases the entropy $\Delta S_{\text{I}}=\ln2$ so
$\Sigma$ has positive heat $\Delta Q_{\text{I}}=T_{0}\ln2$ from $\widetilde
{\Sigma}_{\text{h}}$ and positive work $\Delta W_{\text{I}}=T_{0}\ln2$ on
$\widetilde{\Sigma}_{\text{w}}$ or D, and the first law $\Delta E_{\text{I}%
}=0$ remains satisfied. These quantities are never discussed for the
intervention in the derivation of Landauer's principle. After the gas
equilibrates in $\Sigma$, D picks \textbf{P} in $\widetilde{\Sigma}_{\text{w}%
}$ and\ inserts it at the far right of $\Sigma$. The insertion does not
require any work or heat; see the Insertion Theorem so $\Delta S_{\text{I}%
}^{\prime}=0$ and $\Delta Q_{\text{I}}^{\prime}=0=\Delta W_{\text{I}}^{\prime
}$. Now the gas is isothermally compressed (entropy loss $\Delta S_{\text{E}%
}=-\ln2$) to $V_{1}$ to create the state $0$ during which D or $\widetilde
{\Sigma}_{\text{w}}$\ performs work so $\Delta W_{\text{E}}=-T_{0}\ln2$ \ and
$\widetilde{\Sigma}_{\text{h}}$ receives heat so $\Delta Q_{\text{E}}%
=-T_{0}\ln2$. The particle always ends in the state $0$, regardless of whether
it was initially in the state $0$ or $1$. The entire process is called
"\emph{reset} to $0$," and is considered a logically irreversible operation.
However, net entropy change is $\Delta S=\Delta S_{\text{I}}+\Delta
S_{\text{I}}^{\prime}+\Delta S_{\text{E}}=0$ so the reset is a
thermodynamically reversible operation with no net work done $\Delta W$ and
heat gained $\Delta Q$ by the gas as was the case for Szilard's engine. As the
net heat loss by $\widetilde{\Sigma}_{\text{h}}$ is also zero, there is a
contradiction with the dissipation principle of Landauer. The entire logical
operation is thermodynamically reversible with zero reset cost. The same
reasoning also applies to the general set up with $N$ paritcles with the same
conclusion for inserting \textbf{P }above.

\textbf{Maxwell's Demon}: While it is commonly asserted that Szilard's
information approach is designed to exorcise Maxwell's demon, there is really
no connection between the two. Szilard's pressure demon always remains outside
the system as part of $\widetilde{\Sigma}_{\text{w}}$ and does work on the
system to reduce its entropy without any violation of the second law during
$\Delta\tau_{\text{I}}$. Maxwell's demon is an internal part of $\Sigma$ that
is isolated \cite{Rex}, and is claimed to internally create a temperature
difference $\Delta T$ and a corresponding entropy reduction $\Delta
S_{\text{sort}}$ by sorting and then allowing only slow ($v_{\text{s}}<\bar
{v}$) and fast ($v_{\text{f}}<\bar{v}$) particles across a hole; here $\bar
{v}$ is the average speed of energy $3T_{0}/2$. However, $\Delta
S_{\text{sort}}<0$ will invalidate the second law. The sorting, \textit{i.e.,}
observing fast and slow particles relative to $\bar{v}$ occurs prior to the
intervention of opening the hole to let them through in different parts
$\Sigma_{1}$ and $\Sigma_{2}$ followed by closing it in succession. This order
is reverse of that considered by Szilard and Landauer and causes a major
difference. As D gains information about $\bar{v},v_{\text{s}}$, and
$v_{\text{f}}$\ of incoming particles, there is an information entropy
production $\Delta s_{\text{SL}}$ per observation. It is followed by
interventions only for fast and slow particles. However, the effects of the
two successive parts (opening and closing) of each intervention cancel each
other out so the combined changes are $\Delta S_{\text{I}}=0$ and $\Delta
Q_{\text{I}}=0=\Delta W_{\text{I}}$ per intervention\ (particle transfer).
This scenario does not seem right as D observes particles with $\bar{v}$ lot
more often than those with $v_{\text{s}}$ and $v_{\text{f}}$ so $\Delta
S_{\text{SL}}$ is mostly determined by observing particles with $\bar{v}$ and
will far exceed the entropy reduction $\Delta S_{\text{sort}}$ due to particle
transfers; the idea may seem similar to that proposed by Still \cite{Still}
but is not identical \cite{Note1}. Indeed, our recent investigation
\cite{Gujrati-NEQ-MD}\ shows that the Maxwell's demon cannot destroy
equilibrium so $\Delta T=0\Longrightarrow\Delta S_{\text{sort}}=0$.

In summary, if anything, the current information theories are formulated by
not satisfying the first law during $\Delta\tau_{\text{I}}$, an observation
that has neither ever been made before nor its consequences discussed. Once
$\Delta S_{\text{I}},\Delta Q_{\text{I}}$, and $\Delta W_{\text{I}}$ due to
intervention are included to satisfy the first law, there is no support in the
setup considered for the conjectures by Szilard and Landauer for any
information entropy $\Delta S_{\text{SL}}$. Thermodynamics is sufficient to
yield all the information externally without any need to probe the interior of
the system.


\begin{thebibliography}{99}
\bibitem {Note}It is found useful to keep $\widetilde{\Sigma}_{\text{h}}$ and
$\widetilde{\Sigma}_{\text{w}}$, see Fig. \ref{Fig-Intervention}, separate
from each other as work is an isentropic quantity, while heat is related to
entropy change as they appear in the first law $dE=dQ-dW$. This will help us
distinguish work due to interactions with D and $\widetilde{\Sigma}_{\text{w}%
}$ (blue filled arrow)\ separate from any heat that must be assigned to the
interaction (red filled arrow) with $\widetilde{\Sigma}_{\text{h}}$.
\end{thebibliography}
\end{document}